# Topological basis of signal integration in the transcriptional-regulatory network of the yeast, *Saccharomyces cerevisiae*

Illés J Farkas[1,2]*, Chuang Wu[3]*, Chakra Chennubhotla[3], Ivet Bahar[3], Zoltán N Oltvai[1§]

[1]Department of Pathology, University of Pittsburgh, Pittsburgh, PA, 15261, USA

[2]Department of Biological Physics and HAS Group, Eötvös University, Budapest, 1117, Hungary

[3]Department of Computational Biology, University of Pittsburgh, Pittsburgh, PA, 15261, USA

*These authors contributed equally to this work
§Corresponding author

Email addresses:
    IJF: fij@elte.hu
    CW: chuangwoo@gmail.com
    CC: chakra@ccbb.pitt.edu
    IB: bahar@ccbb.pitt.edu
    ZNO: oltvai@pitt.edu

# Abstract

**Background**

Signal recognition and information processing is a fundamental cellular function, which in part involves comprehensive transcriptional regulatory (TR) mechanisms carried out in response to complex environmental signals in the context of the cell's own internal state. However, the network topological basis of developing such integrated responses remains poorly understood.

**Results**

By studying the TR network of the yeast *Saccharomyces cerevisiae* we show that an intermediate layer of transcription factors naturally segregates into distinct subnetworks. In these topological units transcription factors are densely interlinked in a largely hierarchical manner and respond to external signals by utilizing a fraction of these subnets.

**Conclusions**

As transcriptional regulation represents the 'slow' component of overall information processing, the identified topology suggests a model in which successive waves of transcriptional regulation originating from distinct fractions of the TR network control robust integrated responses to complex stimuli.

# Background

Living cells continuously process information about their environment, and based on this information and their own internal state mount appropriate responses to these signals. This information processing is carried out by various regulatory networks functioning in a highly crowded, viscous cellular interior, with characteristic times spanning several orders of magnitude. The fastest among these are signal transduction networks: they range from simple two-component pathways in prokaryotes to the highly complex signal transduction networks of mammalian cells. Fast signaling, however, is frequently followed by slower transcriptional regulatory (TR) events, during which regulatory gene products, such as transcription factors (TFs) and regulatory RNAs, alter the rate of transcription of other genes, reorganizing gene expression to achieve new metabolic states, or initiate cellular programs, such as the cell cycle, sporulation, or differentiation [1-3].

Understanding the system-level properties of these networks requires both experimental and computational efforts that start with mapping out potential regulatory interactions that exist in a given cell type. In the yeast *Saccharomyces cerevisiae* and in the bacterium *Escherichia coli*, the static 'wiring diagrams' of potential TF-mediated interactions have been mapped out to such a degree [4-7] that their system-level characteristics and function can be investigated. Subsequent computational analyses have shown that in both TR networks the regulatory interactions between TFs and the regulated genes are often organized into basic information processing subgraphs, called motifs [8] that can form even larger potential information processing units, such as motif clusters [9], themes and thematic maps [10], and transcriptional modules [11]. It is also evident that the TR network is utilized in a condition-specific manner [12], perhaps through the activation of distinct, signal-specific subnetworks [13]. In spite of these advances the principles along which regulatory networks process signals, encode the relevant signals at different layers of the network, and integrate them with other signals remain poorly understood.

Here we show that regulatory interactions among an intermediate layer of transcription factors is a key determinant of information transfer within the *S.*

*cerevisiae* TR network, and that this layer naturally segregates into distinct, sparsely communicating subnets in which TFs are densely interlinked in a hierarchical manner. We also show that TFs and the genes regulated by them respond to external signals by utilizing various fractions of these subnetworks. The identified features suggest a model in which successive waves of transcriptional regulation of gene expression via multiple interferences at various levels of TF interaction hierarchy constitute a key feature of developing robust integrated responses to complex stimuli.

## Results

**Hierarchies and signal-specific subnets in the S. cerevisiae TR network**

With the exception of a few mutually regulating pairs, the links of the *S. cerevisiae* TR network are unidirectional, and its nodes can be arranged into three main layers based on their position, regulation, and function. The layers reflect the flow of information from the input nodes (TFs not regulated *transcriptionally* by other TFs), through intermediate TFs to the output nodes (non-TF proteins) (Fig. 1*A*); a path from an input to an output node contains usually 1 to 3 steps, and the maximum length is 8 steps.

In the *S. cerevisiae* TR network each TF regulates a limited number of target genes (intermediate layer TFs and/or output proteins), with an average number of 34.3. As described recently for the TR network of *E. coli* [13], the genes directly or indirectly regulated by a given input TF form a signal-specific subnet, or *origon*, and the nodes at the intermediate and output layers of the origons are often shared by two or more origons. Figure 1*A* illustrates two overlapping origons, originating from the input TFs Yap1 and Skn7. Since the network contains 54 input TFs, there is a total of 54 origons in the *S. cerevisiae* TR network, of which only two are isolated from the rest of the network (the origons of Pdr3 and Zap1) (Fig. 1*B*).

**Classification of the yeast TR network based on its global topological properties**

To gain insight into the overall yeast TR network organization we first assessed the connectivity distribution of all nodes (each representing a gene and its product), and separately those of input TFs, intermediate TFs, and output genes, using cumulated distributions that are equivalent to rank-degree (or Zipf-) plots. Due to the inherent directionality of the links, we separately analyzed the number of regulating TFs per regulated gene (incoming links, $k_{in}$) and the number of regulated genes per TF (outgoing links, $k_{out}$), to determine if their distributions are best approximated by exponential-like [14] or power-law [15] models. (Hubs, i.e., TFs with large numbers of links, are absent from exponential-like models, while they are present and rather significant in the power-law model.) We find that the distribution of the number of incoming links per node, $k_{in}$, displays an exponential decay (see inset of Fig. 1*C*), as previously described [16], while that of outgoing links shows an intermediate behavior between exponential-like- and power-law decay models (Fig. 1*C*).

Interestingly, the outgoing links for *input* TFs closely approximate an exponentially decaying degree distribution, (i.e., hub sizes are limited), while a few of the intermediate TFs are unexpectedly large hubs resembling more closely the power-law models. Also, the outdegrees of intermediate TFs tend to be larger than those of input nodes (Supplementary Fig. S1). Taken together, the cumulative in- and outdegree distributions suggest that the yeast TR network belongs to a mixed class of networks (between exponential and power-law [17]), where the number of connections per node is likely to be constrained both by the limited size of a target gene's promoter region [16], and perhaps by the biosynthetic costs of maintaining regulatory interactions [17].

**Distribution of graph motifs in the yeast TR network**

The effects of many external and internal signals are manifested by altered TF activity, followed by the propagation of the perturbation to nodes of lower layers. Small circuits (or subgraphs) play a key role in this propagation; they often connect nodes of different regulatory layers to each other. Of these, overrepresented subgraphs (motifs) are likely to enhance the versatility of information processing in a TR network [8,18], and may have become abundant due to the overall functional robustness they provide during evolutionary adaptation to changing environmental conditions (see, e.g., Refs. [19-21]).

To elucidate the type and information processing role of such overrepresented subgraphs, we examined the abundance of three-node subgraphs in the *S. cerevisiae* TR network. Using a standard link-randomization algorithm (see Methods) we found that the *feed-forward loop* (FFL), the *single regulatory interaction with mutual regulation* (SMR) and the *convergence with mutual regulation* (CMR) are overrepresented, i.e., they are motifs (Fig. 1*A*), while the *divergence* (DIV), *cascade* (CAS) and *convergence* (CNV) subgraphs are underrepresented, i.e., they are anti-motifs [18] (Table 1). We also examined the position of these 3-node subgraphs with respect to individual origons, and found that (i) similarly to the *E. coli* TR network [13], only a subset of origons contains FFL, SMR, and CMR motifs (Fig. 1*B*), and (ii) the majority (83%) of CNV subgraphs perform signal integration: they receive regulatory signals (directly or indirectly) from two different sources (input TFs) and transmit the joint signal to a single node (Fig. 1*A*).

**Functional cartography of the yeast TR network**

External signals, conveyed by various signaling mechanisms, may be perceived by signal-specific TFs or relatively non-specific TFs. To understand how the responses to these signals are encoded into the topology of the TR network we first examined the degree of overlap among the genes regulated by input- and intermediate TFs. As shown in Figure 2*A* – where the width of a link between two TFs is proportional to the number of outputs (targets) they both regulate – the targets of different TFs extensively overlap (only 3 TFs share no targets with other TFs), suggesting that most genes are combinatorially regulated by several TFs. In contrast, direct regulatory interactions among TFs are more limited (Fig. 2*B*): the largest connected component of the network of direct regulatory interactions among TFs (containing 62 nodes) is sparse, and 30 of the remaining 37 TFs have no regulatory interactions with other TFs at all, i.e., they act in isolation.

To characterize the type of combinatorial regulation performed by each TF, we color coded each of the 99 TFs according to the function(s) of the genes they regulate. To this end, we resorted to the 33 GO Slim biological process terms [22], which we grouped into eight GO Slim categories described in the Methods. It is evident, that all TFs regulate genes with various functions (Fig. 2B). For example, genes within two overlapping origons – defined by the input TFs Ino4 and Stb1 – display a multitude of functions (Fig. 2*C*). Stb1 takes part in the regulation of transcription at the G1/S transition [23], while Ino4 is a positive regulator of phospholipid biosynthesis [24].

Similarly to Stb1, the two intermediate TFs, Swi5 and Ndd1, regulate temporal expression patterns: Ndd1 is essential for the activation of many late S-phase specific genes [25], while Swi5 activates genes in the G1 phase and at the M/G1 boundary [26]. Notably, in the overlap of the origons Ino4 and Stb1 two major regulatory tasks are integrated (Fig. 2*C*). Among the genes contained exclusively by the Ino4 origon participation in metabolism is very common, while only one gene is known to perform a cell-cycle related function. For genes contained exclusively by origon Stb1 this relation is reversed, while in the overlap of the two origons both functions are common. Thus, the overlap of these two origons illustrates the coordination of a temporally regulated event (cell cycle) with another general task (phospholipid metabolism).

For a concise analysis of regulatory task integration by overlapping origons, in each of the 418 overlapping origon pairs (A, B), we listed the GO Slim biological process terms for the regions A^B (overlap), A\B and B\A (genes contained exclusively by origon A or B). We found that the distribution of GO Slim biological processes in the set A^B is in general significantly similar (average Z score: 2.2) to the distribution deduced from the sets A\B and B\A summed together (see Methods for details). Thus, we infer that in the TR network of *S. cerevisiae* overlapping pairs of origons significantly integrate regulatory tasks.

**Topological organization of signal integration in the yeast TR network**

Complex environmental signals are decomposed into more elementary signals that eventually elicit an integrated transcriptional response in the context of the cell's own internal state. Since intermediate TFs (by definition) transmit signals from input to output nodes and provide connections among all TFs (Fig. 1*A*), the topological organization of their interactions is likely to play a key role in developing such integrated responses. To examine their relationships, we decomposed the TR network by an iterative peeling algorithm (see Methods), where the top and bottom layers of the network have been successively removed until only 3 small isolated graph components ('cores') remained. Then these cores were consolidated by adding back their nearest up- and downstream intermediate regulators (Fig. 3*A*). After this decomposition procedure we found that the 45-node intermediate TF subnetwork naturally segregated into three internally densely-connected groups of TFs (referred to as 'organizer' O1, O2, and O3 hereafter), as well as several isolated TF nodes (Figs. 3*A,B*). In contrast, the connections between organizers are sparse (Fig. 3*B*): organizers O1 and O2 are connected by one interaction (between Nrg1 and Hap4), and O2 and O3 have only two connections (Fkh1-Yhp1 and Abf1-Put3). Of note, all three inter-organizer connections transfer a signal from the 'top' (as defined by the flow of information) of one organizer to the 'bottom' of the other. We also find that input TFs often co-regulate intermediate TFs located in one or two organizers, but never in all three of them. Note, that as an alternative approach we also performed computational search for partially overlapping communities [27] in the TR network. This analysis yielded highly similar results (Supplementary Fig. S3), suggesting that the concept of organizers is valid irrespective of data stringency (Supplementary Fig. S4), or the analytical technique used for their identification.

Currently, on the global scale the dynamical utilization of signal-specific transcription regulatory subnets can be best tested with microarray expression data [12,13]. To analyze the dynamical role of organizers, for each of the 45 intermediate TFs we have defined the TF and the list of its targets as a group of genes, and computed the transcriptional response of this group to a given external or internal signal (see Methods). Under hyperosmotic shock (Fig. 3*C*), the TFs (and their target genes) in organizer O2 displayed by far the strongest average response, as measured by the double Z score [13] (see Methods): 0.8, compared to -0.13 and -0.14 in organizers O1 and O3, respectively. Within this group the set of genes regulated by intermediate TFs Hap4, Sok2, Phd1, and Rox 1 show the strongest response. All these TFs are regulated by input TF, Skn7, suggesting that this input TF is one of the main sensors of hyperosmotic shock in *S. cerevisiae*, in agreement with previous results [28]. A similar conclusion can be drawn for all other environmental stimuli tested (Supplementary Fig. S5), suggesting that only a subnet of organizer(s) are activated upon simple or complex environmental stimuli.

## Discussion

The multitude of cellular tasks makes it necessary for cellular components to be hierarchically organized into groups based on functional association [29]. One well-studied aspect of this functional organization is the 'static map' of a TR network, i.e., the list of all possible transcription regulatory (TR) interactions within a cell. Small numbers of individual TR nodes (TFs and their regulated genes) are known to be arranged into overrepresented, specifically wired information processing units (motifs) [8], which in turn participate in a series of sequentially embedded higher order structures [9,10]. In an actual response, however, from all topological (static) possibilities in the TR network the cell utilizes only limited sets of these interactions [12]. These interactions are often signal-specific [13], though there are also many TR nodes that are known to be generic responders [12].

However, TR interactions represent only a subset of regulatory interactions. In fact, protein-protein- and protein-metabolite interactions represent the majority of information processing interactions of a cell (Fig. 4). When taking this into account, additional heterogeneous interaction patterns can be uncovered at various hierarchical scales [10,30]. Nevertheless, TR interactions represent the 'slow component' of the overall network, whose behavior determines long-range response [1-3]. Thus, it is of great importance to understand how the large-scale structure of a TR network reflects the integration of the vast variety of individual external signals with each other and with the cell's internal state.

Detailed methods, a supplementary table and supplementary figures are also available [see Additional file 1].

## Conclusions

From the analyses presented here the system-level picture arising for the integration of TR signals in yeast suggests the presence of a small number of large-scale signal integration 'pools' (organizers) along which signals are processed and transmitted towards all target genes (Fig. 4). Regulatory connections inside organizers are dense, while inter-organizer connections are sparse. In addition to this topological separation, the target genes of different organizers also elicit remarkably different transcriptional responses (Fig. 3*C*). Moreover, due to the slowness of the interactions (minute-scale delays due to transcription and translation) a given signal can elicit subsequent waves of transcriptional regulatory events that are usually integrated through feedbacks of rapid interactions (Fig. 4). For example, transcriptional regulation in response to decreasing concentration of oxygen (as Signal X in Fig. 4) is carried out mainly by two TFs, FNR and ArcA in *E. coli*. Although ArcA can be transcriptionally activated by FNR (i.e., ArcA is an intermediate TF), FNR is conformationally activated at a lower oxygen level than ArcA. Thus, ArcA-specific genes are activated first, followed by a subsequent wave of activation of a second set of genes (many co-activated by FNR and ArcA) that partially overlaps with genes activated during the first wave [31,32]. In turn, rapid non-transcriptional feedback, such as phosporylation of TFs, may alter the activity of other intermediate TFs. This may initiate additional sets of 'transcriptional waves' leading to the comprehensive response of the cell observed upon the aerobic-anaerobic shift (Fig. 4).

What explains the evolution of the observed topological structure? The TF network appears to grow by node duplication [33], resulting in structurally related TF protein families, in which diversification is both a result of TF structural evolution [34] and the evolution of DNA binding motifs [35]. The subsequent natural selection of motifs and higher order structures might have been driven by their ability to provide reliable information processing functions to the cell, including robustness against mutations [36], noise [19,20], and oscillating signals [37,38], while simultaneously allowing rapid response to common signals in an overall highly variable environment [21]. The future availability of additional types of interaction maps, such as those of phosphoproteins [39], together with an improved understanding of the behavior of fast- (signaling), slow- (transcriptional) and combined circuits [38,40-42] will probably further explain the emergence of the observed small and large-scale topological structures of the cell's information processing network.

## Methods

**Databases and Software**

The publicly available dataset on the TR network of *Saccharomyces cerevisiae* was downloaded from the supporting website of the original publication [6]. This computationally filtered dataset, originally obtained in rich media and a few other growth conditions, lists directed binary interactions at various confidence levels, and is further improved by including additional transcriptional interactions from the literature [6]. All computational analyses were performed with the SGD IDs of the genes that were then transformed back to traditional gene names for easier presentation. Conversion tables were downloaded from the Saccharomyces Genome Database (SGD) and the MIPS Comprehensive Yeast Genome Database (CYGD). Of the six different datasets representing various confidence levels [6], we used the highest confidence data set for most of our analyses (Supplementary Table S1). Originally, the network derived from this dataset contained 1905 nodes and 3406 regulatory interactions, which we reduced to 1905 nodes and 3394 directed links by removing 12 autoregulatory links. The resulting network contained 99 TFs (54 input and 45 intermediate nodes) and except for two small isolated groups – with the input nodes Pdr3 (drug resistance, regulating itself and one other gene) and Zap1 (zinc-regulated, regulating four other genes) - it is comprised of one giant connected component. Most targets (intermediate and output nodes) are regulated by more than one (on the average, 1.8) TFs. We quantify the relative overlap between the target lists ($A_i$ and $A_j$) of two TFs (*i* and *j*) by the Jaccard correlation, $|A_i \cap A_j|/|A_i \cup A_j|$, between the two sets. An alternative representation of the TR network is to consider only TFs and the regulatory interactions between them, in which case the network contains 99 nodes of which 69 are connected in a giant component.

The normalized microarray expression data sets GDS18-20, GDS112-115, and GDS362 were downloaded from the FTP directory of NCBI's Gene

Expression Omnibus (GEO). Our programs were written in Perl and C++, and for visualization we used the Linux tools Xfig and Gnuplot together with the network drawing program Pajek [43].

**Network randomization and graph motifs**

To assess the enrichment of 3-node subgraphs in the regulatory network, we used link randomization tests [8] that preserve the number of incoming and outgoing links around each node, but obliterate all other information about the connectivity of the network. In one step of this method two links, A→B and C→D, are selected randomly and moved to the unoccupied A→D and C→B positions. We examined $n_N$ = 100 randomized networks, each produced with $n_S$ = 100,000 rewiring steps starting from the original TR network, i.e., each link was moved approximately 60 times to generate a given randomized network. Following Ref. [8] a subgraph with $M_0$ copies in the original TR network and $M±\Delta M$ copies in the randomized versions is called a graph *motif*, provided that the associated Z score, $Z = (M_0 - M) / \Delta M$, is significantly positive. We also verified that for the TR network studied here $n_N$ and $n_S$ are both sufficiently large to ensure the convergence of the Z-scores for 3-node subgraphs.

**Cumulative GO categories**

For functional characterization of yeast proteins we grouped the 33 Gene Ontology (GO) Slim Biological Process terms [22] into the following eight categories: *cell cycle*-related (GO terms: cell cycle, cell budding, conjugation, cytokinesis, meiosis, pseudohyphal growth, sporulation), *metabolism*-related (GO terms: amino acid and derivative metabolism, carbohydrate metabolism, cellular respiration, DNA metabolism, generation of precursor metabolites and energy, lipid metabolism, protein catabolism, RNA metabolism, vitamin metabolism), *morphogenesis*-related (GO terms: cell wall organization and biogenesis, cytoskeleton organization and biogenesis, membrane organization and biogenesis, morphogenesis, nuclear organization and biogenesis, organelle organization and biogenesis, ribosome biogenesis and assembly), *transcription and protein synthesis*-related (GO terms: protein biosynthesis, protein modification, transcription), *transport*-related (GO terms:

electron transport, transport, vesicle-mediated transport), *stress and homeostasis*-related (GO terms: cell homeostasis, response to stress, signal transduction), *cell movement*-related (GO terms: substrate-bound cell migration and cell extension), *unknown* (biological_process, biological_process unknown, unknown), respectively.

**Task integration by overlapping origons**

A simplifying view of the TR network is provided by the origon representation [13], shown by color-coded circles in Figure 1*B*. Each origon represents a cluster of nodes originating from a common (input) TF (54 of them in the present case), and the color code therein describes the occurrence of four types of interaction motifs distinguished by their high Z-scores (see below). Except for the two input nodes mentioned above (Prd3 and Zap1), all origons are interconnected due to the partial overlaps between their members at intermediate and output layers. The number of shared members is reflected by the thickness of the links between the origons. The examined yeast TR network has 418 such overlapping pairs of origons.

Of interest is to characterize the degree of integration of functional tasks between overlapping pairs of origons. To this aim, we first removed from the TR network all gene (products) with GO Slim annotation "unknown", and counted the number of genes annotated by a given GO Slim term, within the subsets A^B (overlap), A\B and B\A (genes contained only by A or B) for each pair of overlapping origons (A. B).Three vectors, defined by the fractions/probabilities of GO Slim terms were thus generated for each pair, denoted as ***a*** (for A\B), ***b*** (for B\A), or ***c*** (for A^B). The overlap (A^B) integrates tasks from the other two regions, if ***c*** is sufficiently similar to both ***a*** and ***b***. The extent of similarity between the three probability distributions was then assessed by the correlation cosines (***c . a***) and (***c . b***), expressed by the sum K = ***c . (a+b)***, where the dot designates the scalar product. We found that the K values for pairs of origons in the yeast TR network were significantly higher than those calculated for 100 randomized test cases. The corresponding Z score – i.e. (<original K value>–<average K in random cases>) /<standard deviation in random cases> – averaged over all pairs was <Z(K) > = 2.2.

**Locating densely connected subnetworks (organizers) of Transcription Factors**

In the network of TFs (nodes: Transcription Factors, links: regulatory interactions) we identified subnetworks distinguished by their dense interconnection and central role (i.e., organizers) by using an iterative layer-peeling algorithm [44], as follows. After first removing all autoregulatory loops, we repeatedly removed the nodes in the top and bottom layers of the network until only three small isolated (graph) components ('cores') remained. To these cores we then added in 3 subsequent steps their up- and downstream intermediate regulators to obtain three major organizers (see Results).

Alternatively, to locate overlapping, densely connected groups of nodes among the 69 non-isolated TFs we applied CFinder [45] to the underlying undirected network and identified the *k*-clique communities (groups of densely interconnected nodes) at *k* = 3 corresponding to 'rolling' a triangle by moving one of its nodes at each step.. Note that any TF (node) was allowed to belong to more than one community. Next, we added to each community, $C_A$, all nodes reachable from a node of $C_A$ via regulatory interactions, but not yet contained by any of the communities. Last, we merged communities $C_A$ and $C_B$, if all exclusively contained nodes of $C_A$ were directly regulated by an exclusively contained node of $C_B$.

**Significance of the transcriptional response of a group of genes**

Our goal was to quantify the effect of particular (environmental or internal) conditions (or signals) *S* on the transcript levels of a selected group of genes. First, we grouped experiments (GSMs, Geo SaMples) according to their platforms (GPLs). Then to each experiment obtained under a 'normal' condition (*e.g.*, stationary state) we assigned the signal *S=-1* and to all others (*e.g.*, hyper-osmotic shock, N depletion, or DNA damage with MMS) we assigned the signal *S=+1*. Next, we computed the Pearson correlation, $C_i$, between the *i*th gene's expression $E_{ij}$ and the *j*th experimental condition $S_j$. using

$$C_i(E_{ij}, S_j) = \frac{\langle E_{ij} S_j \rangle_j - \langle E_{ij} \rangle_j \langle S_j \rangle}{\left[\langle E_{ij}^2 \rangle_j - \langle E_{ij} \rangle_j^2\right]^{1/2} \left[1 - \langle S_j \rangle^2\right]^{1/2}},$$

where the subscript *j* includes both those experiments under the condition of interest (i.e. experiments $a_1$, $a_2$, …, $a_n$, signal value: $S_j=+1$) and those under 'normal' conditions ($j=b_1$, $b_2$, …, $b_m$, and $S_j=-1$). The $i$th gene's response to signal $S$ is significant, *i.e.*, it is strongly activated (repressed), if its $C_i$ value is higher (lower) than the majority of the correlation values calculated for all yeast genes. This can be measured with the Z score, $Z_i = |C_i - C|/\Delta C$, of the $i$th gene's response, where $C$ and $\Delta C$ are the average and standard deviation of the correlation values of all yeast genes. Here we use the absolute value, because a strong activation and a strong repression are equally important responses and should both give a high Z score.

The significance of the response of the entire group $G$ to condition $S$ can be assessed by comparing the average Z score in $G$, $Z_G = \langle Z_i \rangle_{i \in G}$, to the similarly computed averages ($Z_{H1}$, $Z_{H2}$, …) in other, randomly selected groups of genes of the same size ($H1$, $H2$, …). We used 1,000 such control groups. Denoting by $\langle Z_H \rangle$ and $\Delta Z_H$ the average and standard deviation of $Z_H$ values, the double Z score of the response of group $G$ is $Y_G = (Z_G - \langle Z_H \rangle)/\Delta Z_H$.

## Authors' contributions

Computational analyses were carried out by IJF, CW and CC, figures were produced by IJF and CW, with direction from ZNO, IB and CC. The manuscript was written by IJF, CW, IB and ZNO and edited by all authors.

## Acknowledgements

We thank G. Balázsi and T. Vicsek for discussion and comments on the manuscript. IB gratefully acknowledges support from NIH Award # P20 GM065805-02. Research by IJF at Eötvös University was supported by the Hungarian Scientific Research Fund (OTKA, Grants No. D048422 and F047203).

# References

1. Babu MM, Luscombe NM, Aravind L, Gerstein M, Teichmann SA: **Structure and evolution of transcriptional regulatory networks.** *Curr Opin Struct Biol* 2004, 14: 283-291
2. Blais A, Dynlacht BD: **Constructing transcriptional regulatory networks.** *Genes Dev* 2005, 19: 1499-1511
3. Davidson EH, Erwin DH: **Gene regulatory networks and the evolution of animal body plans.** *Science* 2006, 311: 796-800
4. Svetlov VV, Cooper TG: **Review: compilation and characteristics of dedicated transcription factors in Saccharomyces cerevisiae.** *Yeast* 1995, 11: 1439-1484
5. Lee TI, Rinaldi NJ, Robert F, Odom DT, Bar-Joseph Z, Gerber GK, Hannett NM, Harbison CT, Thompson CM, Simon I, Zeitlinger J, Jennings EG, Murray HL, Gordon DB, Ren B, Wyrick JJ, Tagne JB, Volkert TL, Fraenkel E, Gifford DK, Young RA: **Transcriptional regulatory networks in Saccharomyces cerevisiae.** *Science* 2002, 298: 799-804
6. Harbison CT, Gordon DB, Lee TI, Rinaldi NJ, Macisaac KD, Danford TW, Hannett NM, Tagne JB, Reynolds DB, Yoo J, Jennings EG, Zeitlinger J, Pokholok DK, Kellis M, Rolfe PA, Takusagawa KT, Lander ES, Gifford DK, Fraenkel E, Young RA: **Transcriptional regulatory code of a eukaryotic genome.** *Nature* 2004, 431: 99-104
7. Salgado H, Santos-Zavaleta A, Gama-Castro S, Peralta-Gil M, Penaloza-Spinola MI, Martinez-Antonio A, Karp PD, Collado-Vides J: **The comprehensive updated regulatory network of Escherichia coli K-12.** *BMC Bioinformatics* 2006, 7: 5
8. Milo R, Shen-Orr S, Itzkovitz S, Kashtan N, Chklovskii D, Alon U: **Network motifs: simple building blocks of complex networks.** Science 2002, 298: 824-827
9. Dobrin R, Beg QK, Barabasi AL, Oltvai ZN: **Aggregation of topological motifs in the Escherichia coli transcriptional regulatory network.** *BMC Bioinformatics* 2004, 5: 10
10. Zhang LV, King OD, Wong SL, Goldberg DS, Tong AH, Lesage G, Andrews B, Bussey H, Boone C, Roth FP: **Motifs, themes and thematic maps of an integrated Saccharomyces cerevisiae interaction network.** *J Biol.* 2005, 4: 6
11. Resendis-Antonio O, Freyre-Gonzalez JA, Menchaca-Mendez R, Gutierrez-Rios RM, Martinez-Antonio A, Avila-Sanchez C, Collado-Vides J: **Modular analysis of the transcriptional regulatory network of E. coli.** *Trends Genet.* 2005, 21: 16-20
12. Luscombe NM, Babu MM, Yu H, Snyder M, Teichmann SA, Gerstein M: **Genomic analysis of regulatory network dynamics reveals large topological changes.** *Nature* 2004, 431: 308-312
13. Balazsi G, Barabasi AL, Oltvai ZN: **Topological units of environmental signal processing in the transcriptional regulatory network of Escherichia coli.** *Proc Natl Acad Sci U S A* 2005, 102: 7841-7846
14. Erdős P, Rényi A: **On the evolution of random graphs.** *Publ. Math. Inst. Hung. Acad. Sci.* 1960, 5: 17-61

15. Barabasi AL, Albert R: **Emergence of scaling in random networks.** *Science* 1999, 286: 509-512
16. Guelzim N, Bottani S, Bourgine P, Kepes F: **Topological and causal structure of the yeast transcriptional regulatory network.** *Nat Genet.* 2002, 31: 60-63
17. Amaral LA, Scala A, Barthelemy M, Stanley HE: **Classes of small-world networks.** *Proc Natl Acad Sci U S A.* 2000, 97: 11149-11152
18. Milo R, Itzkovitz S, Kashtan N, Levitt R, Shen-Orr S, Ayzenshtat I, Sheffer M, Alon U: **Superfamilies of evolved and designed networks.** *Science* 2004, 303: 1538-1542
19. Klemm K, Bornholdt S: **Topology of biological networks and reliability of information processing.** *Proc Natl Acad Sci U S A.* 2005, 102: 18414-18419
20. Prill RJ, Iglesias PA, Levchenko A: **Dynamic properties of network motifs contribute to biological network organization.** *PLoS Biol.* 2005, 3: e343
21. Kashtan N, Alon U: **Spontaneous evolution of modularity and network motifs.** *Proc Natl Acad Sci U S A.* 2005, 102: 13773-13778
22. Martin D, Brun C, Remy E, Mouren P, Thieffry D, Jacq B: **GOToolBox: functional analysis of gene datasets based on Gene Ontology.** *Genome Biol.* 2004, 5: R101
23. Ho Y, Costanzo M, Moore L, Kobayashi R, Andrews BJ: **Regulation of transcription at the Saccharomyces cerevisiae start transition by Stb1, a Swi6-binding protein.** *Mol Cell Biol.* 1999, 19: 5267-5278
24. Ambroziak J, Henry SA: **INO2 and INO4 gene products, positive regulators of phospholipid biosynthesis in Saccharomyces cerevisiae, form a complex that binds to the INO1 promoter.** *J Biol Chem.* 1994, 269:15344-15349
25. Loy CJ, Lydall D, Surana U: **NDD1, a high-dosage suppressor of cdc28-1N, is essential for expression of a subset of late-S-phase-specific genes in Saccharomyces cerevisiae.** *Mol Cell Biol.* 1999, 19: 3312-3327
26. McBride HJ, Yu Y, Stillman DJ: **Distinct regions of the Swi5 and Ace2 transcription factors are required for specific gene activation.** *J Biol Chem.* 1999, 274: 21029-21036
27. Palla G, Derenyi I, Farkas I, Vicsek T: **Uncovering the overlapping community structure of complex networks in nature and society.** *Nature* 2005, 435: 814-818
28. Garay-Arroyo A, Lledias F, Hansberg W, Covarrubias AA: **Cu,Zn-superoxide dismutase of Saccharomyces cerevisiae is required for resistance to hyperosmosis.** *FEBS Lett.* 2003, 539: 68-72
29. Papin JA, Reed JL, Palsson BO: **Hierarchical thinking in network biology: the unbiased modularization of biochemical networks.** *Trends Biochem Sci.* 2004, 29: 641-647
30. Yeger-Lotem E, Sattath S, Kashtan N, Itzkovitz S, Milo R, Pinter RY, Alon U, Margalit H: **Network motifs in integrated cellular networks of transcription-regulation and protein-protein interaction.** *Proc Natl Acad Sci U S A.* 2004, 101: 5934-5939

31. Kiley PJ, Beinert H: **Oxygen sensing by the global regulator, FNR: the role of the iron-sulfur cluster.** *FEMS Microbiol Rev.* 1998, 22: 341-352
32. Sawers G: **The aerobic/anaerobic interface.** *Curr Opin Microbiol.* 1999, 2: 181-187
33. Teichmann SA, Babu MM: **Gene regulatory network growth by duplication.** *Nat Genet.* 2004, 36: 492-496
34. Babu MM, Teichmann SA: **Evolution of transcription factors and the gene regulatory network in Escherichia coli.** *Nucleic Acids Research* 2003, 31: 1234-1244
35. Ihmels J, Bergmann S, Gerami-Nejad M, Yanai I, McClellan M, Berman J, Barkai N: **Rewiring of the yeast transcriptional network through the evolution of motif usage.** *Science* 2005, 309: 938-940
36. Kafri R, Bar-Even A, Pilpel Y: **Transcription control reprogramming in genetic backup circuits.** *Nat Genet.* 2005, 37: 295-299
37. Lipan O, Wong WH: **The use of oscillatory signals in the study of genetic networks.** *Proc Natl Acad Sci U S A* 2005, 102: 7063-7068
38. Pomerening JR, Kim SY, Ferrell JE: **Systems-level dissection of the cell-cycle oscillator: bypassing positive feedback produces damped oscillations.** *Cell* 2005, 122: 565-578
39. Ptacek J, Devgan G, Michaud G, Zhu H, Zhu X, Fasolo J, Guo H, Jona G, Breitkreutz A, Sopko R, McCartney RR, Schmidt MC, Rachidi N, Lee SJ, Mah AS, Meng L, Stark MJ, Stern DF, De Virgilio C, Tyers M, Andrews B, Gerstein M, Schweitzer B, Predki PF, Snyder M: **Global analysis of protein phosphorylation in yeast.** *Nature* 2005, 438: 679-684
40. Mangan S, Zaslaver A, Alon U: **The coherent feedforward loop serves as a sign-sensitive delay element in transcription networks.** *J Mol Biol.* 2003, 334: 197-204
41. Brandman O, Ferrell JE Jr, Li R, Meyer T: **Interlinked fast and slow positive feedback loops drive reliable cell decisions.** *Science* 2005, 310: 496-498
42. Guido NJ, Wang X, Adalsteinsson D, McMillen D, Hasty J, Cantor CR, Elston TC, Collins JJ: **A bottom-up approach to gene regulation.** *Nature* 2006, 439: 856-860
43. Batagelj V, Mrvar A: **PAJEK -- Program for large network analysis.** *Connections* 1998, 21: 47-57
44. Wuchty S, Almaas E: **Peeling the yeast protein network.** *Proteomics* 2005, 5: 444-449
45. Adamcsek B, Palla G, Farkas IJ, Derenyi I, Vicsek T: **CFinder: locating cliques and overlapping modules in biological networks.** *Bioinformatics* 2006, 22: 1021-1023
46. Ma'ayan A, Jenkins SL, Neves S, Hasseldine A, Grace E, Dubin-Thaler B, Eungdamrong NJ, Weng G, Ram PT, Rice JJ, Kershenbaum A, Stolovitzky GA, Blitzer RD, Iyengar R: **Formation of regulatory patterns during signal propagation in a Mammalian cellular network.** *Science* 2005, 309: 1078-1083

# Figures

**Figure 1 - Global organization of the yeast transcriptional-regulatory network**

(**A**) The hierarchical arrangement of the TR network into input, intermediate and output layers (rectangles, ellipses, and small circles, separated by dashed lines, respectively) is illustrated for two partially overlapping origons, Yap1 and Skn7. The boxes illustrate 3-node subgraphs, CNV, CMR, and FFL distinguished by their high frequency of occurrence in the yeast TR network (Table 1). (**B**) The network of origons [13] in the *S. cerevisiae* TR network. Each circle represents an origon labeled by its input TF. The size of each circle is proportional to the number of genes in that origon. Two origons are connected if they share at least one gene and the width of a link is proportional to the number of genes that the two connected origons share. Three different types of subgraphs, indicated by the colored labels are distinguished in the origons (see Table 1). The fractional area of each color on the origon circle is proportional to the number of occurrences of the corresponding subgraph among the members of the origon. If an origon contains none of the listed subgraphs, it is shown in grey color. (**C**) Main panel: the distribution of outdegrees (number of outgoing connections of a node, $k_{out}$) shows that this network falls between models with an exponential or faster degree distribution cutoff [14,17] and the scale-free model [15] (with some difference for input and intermediate TF nodes), though neither of the two types of models is significantly closer than the other.

**Figure 2 - Signal integration in the yeast TR network**

(**A**) The network of input (brown) and intermediate (purple) TFs is shown. The size of a node is proportional to the number of genes it regulates, while the width of a line connecting two nodes is proportional to the number of target genes jointly regulated by the two TFs. Except for Pdr3, Zap1 (input TFs) and Mot2 (intermediate TF), TFs are strongly connected to each other (i.e., share many of their target genes), indicating that the functions of the TFs are widely integrated, and that most genes are jointly or combinatorially regulated by groups of regulators, rather than individual ones. (**B**) Functional cartography in the network of TFs. Each node represents one TF and each link represents a regulatory interaction. The area of a TF node is proportional to the number of genes it regulates, and colors refer to the GO Slim annotation distributions of its target genes (see Methods for details). Input nodes are encircled by thick black lines. Regulatory links from input to intermediate TFs are shown in black, while links among intermediate TFs are colored red. The single unidirectional cycle connecting Dig1, Tec1 and Ste12 is shown by thick red edges. The portion enclosed in the dashed box is enlarged in panel C. (**C**) The overlapping origons Ino4 and Stb1 integrate cellular functions (see text for detailed analysis). Enlarged versions of panels A-C are provided as Supplementary Figure S2A-C.

**Figure 3 - Topological organization of signal integration**

(**A**) Decomposition of the TR network by removing its top- (input TFs) and bottom layers (output nodes) identifies the intermediate TF layer, which, based on the high local density and distribution of connections, is naturally subdivided into three major groups (organizers), as well as a number of isolated TFs. The connections between organizers are sparse. Nodes are arranged hierarchically based on the direction of information flow. The chain of links colored red shows the longest path through the network. Regulatory signals flow from darker nodes towards lighter ones. (**B**) The three emerging organizers of the yeast TR network are enclosed by blue, red, and green rectangles, respectively, while isolated intermediate TFs are on the right. The relative size and color code of each node conform to the descriptions given in Fig. 2*B*. Within organizers the density of links is more than 10 times higher than that between the organizers. Input TF nodes regulating the intermediate TFs in the organizers are shown by rectangles. The blue nodes on the left side of O1, the green ones on the right of O3, and the red ones above/below O2 are the inputs that regulate each one organizer. The magenta, cyan and yellow nodes regulate pairs of organizers, as indicated by the links. Note that there is no input TF regulating all the three organizers. The number of transcriptional inputs for each of the intermediate TFs is shown in parentheses. Essential TFs (+) and those with autoregulatory loops (@) are indicated. (**C**) Transcriptional response of organizers to hyperosmotic shock.

The double Z scores (ordinate) [13] measure the significance of the response of each organizer node plus its target genes to the external condition as compared to the control condition (a strong up- and downregulation both give a high Z score). The numbers in the bottom part of each graph denote the average double Z scores for O1 (blue) O2 (red) and O3 (green), respectively, while the colored dots represent the average double Z-score of genes regulated by the indicated intermediate TF. Black dots represent the same for the input TF(s) directly regulating the indicated intermediate TF.

**Figure 4 - Schematic representation of intracellular information processing**

Figure legend text. The transcriptional-regulatory (TR) network is composed of input TFs (not regulated by other TFs) (squares), intermediate TFs (regulated by at least one other TF) (circles) and output nodes (regulated effector genes) (triangles). Signals external to the cell can affect the input- and at least some of the intermediate TFs directly or indirectly through signaling cascades. Internal signals, through the activity of the overall molecular interaction network of the cell (shaded in grey) can potentially affect all nodes of the TR network through allosteric regulation, posttranslational modification, etc. Within the TR network the various signals are integrated within relatively distinct subnetworks, or organizers (brown-shaded boxes) composed of intermediate TFs. The TFs within organizers are densely linked but there are only sparse links with TFs in other organizers. A given elementary signal (e.g., Signal X) may affect only a single origon [13], depicted here as the filled symbols, but complex signals may affect several origons simultaneously. As transcription is the 'slow' component of the overall regulatory network in which each link adds a time delay in the regulation, there is a very rich possibility of dynamics carried out on the topology. In particular, nodes might be activated at several time steps (represented by the different fill patterns) corresponding to the propagation of subsequent reaction waves in chemical/interaction space [46].

# Tables

**Table 1 - Number distributions and statistical significance of 3-node subgraphs in yeast TR network**

| Subgraph | 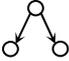 DIV (divergence) | 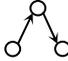 CAS (cascade) | 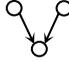 CNV (convergence) | 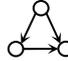 FFL (feed-fwd loop) | 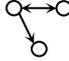 SMR (single link with mutual regulation) | 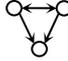 CMR (convergence with mutual regulation) |
|---|---|---|---|---|---|---|
| Number in the original network | 150 845 | 2 898 | 2 655 | 392 | 307 | 118 |
| After link randomization | 151 477 ± 152 | 3 543 ± 156 | 2 996 ± 23 | 176 ± 22 | 126 ± 148 | 2.6 ± 3.9 |
| Significance of original (Z score) | *-4.2* | *-4.1* | *-15* | **<u>9.7</u>** | **<u>1.2</u>** | **<u>30</u>** |
| Subgraph type | *Anti-motif* | *Anti-motif* | *Anti-motif* | **Motif** | **Motif** | **Motif** |

Motifs are marked, and only subgraphs with at least 100 occurrences in the original network are listed. After link randomization the numbers of FFL, SMR and CMR subgraphs decrease, while those of DIV, CAS, and CNV subgraphs are maintained with slight increases, indicating that FFL, SMR and CMR are motifs in the TR network, while DIV, CAS and CNV are anti-motifs [8,18].

# Additional files

**Additional file 1 – Supplementary Material**
Format: PDF.
Contents: Detailed Methods, Supplementary Table, Supplementary Figures.